\documentclass[twocolumn,showpacs,preprintnumbers,amsmath,amssymb,prb]{revtex4}
\usepackage{graphicx}
\usepackage{dcolumn}
\usepackage{bm}
\usepackage{color}

\newcommand{\cblue}[1]{{\color{blue} #1}}

	\preprint{APS/123-QED}	

\begin{document}
	
	\title{		Indication 
	 of novel magnetoresistance mechanism 
	in (Bi,Sb)$_2$(Te,Se)$_3$ 3D topological insulator thin films
	}
	\author{N.~P.~Stepina$^{1*}$, A.~O.~Bazhenov$^1$, A.~V.~Shumilin$^{2,3}$, A.~Yu.~Kuntsevich$^4$, V.~V. ~Kirienko$^1$, E.~S.~Zhdanov$^1$,   D.~V.~Ishchenko$^1$, O.~E.~Tereshchenko$^{1}$}
\email{nstepina@mail.ru}
\affiliation{$^1$Institute of Semiconductor Physics,  630090 Novosibirsk, Russia}
\affiliation{$^2$Ioffe Institute, 194021 St.-Petersburg, Russia}
\affiliation{$^3$Jozef Stefan Institute, Ljubljana, 1000, Slovenia}
\affiliation{$^4$P. N. Lebedev Physical Institute, Russian Academy of Sciences, 119991 Moscow , Russia}

\begin{abstract}
Electron states with the spin-momentum-locked Dirac dispersion  at the surface of a three-dimensional (3D) topological insulator are known to
lead to weak antilocalization (WAL), i.e. low temperature and low-magnetic field quantum interference-induced  positive magnetoresistance (MR).
In this work we report on the MR measurements in  (Bi,Sb)$_2$(Te,Se)$_3$ 3D topological insulator thin films epitaxially grown on Si(111), demonstrating an anomalous WAL amplitude. 
This anomalously high amplitude of WAL 
can not be explained by parabolic or linear MR and indicates the existence of an additional, 
MR mechanism. Another supporting observation is not linear in the classically weak magnetic field 
Hall effect in the same films.
The increase of the low-field Hall coefficient, with respect to the higher-field  value, reaches 
 10$\%$. 
We 
 consistently explain both transport features within a two-liquid model, where the mobility of one of the components drops strongly in a weak magnetic field.  We argue that this dependence may arise from 
the Zeeman field induced gap opening mechanism. 

\end{abstract}

	\maketitle
	
	\section{Introduction}
 
	Magnetotransport in a three-dimensional (3D) topological insulators (TIs) has been a matter of intensive investigations
	in recent years~\cite{Tian, Lang, Liu2011}. Layered 3D TIs of tetradymite structure (Bi$_2$Se$_3$, Bi$_2$Te$_3$ and so on) are the most studied~\cite{AndoJPSJ,Li_Zhang}. The conductivity of films and crystals of these materials is generally believed to consist of three channels connected in parallel: bulk and two surfaces (top and bottom). Reveling the topologically protected surface states (TSSs) requires the suppression of  bulk conductivity by  reducing the thickness of the film or flake. 
	This three-channel picture has become paradigmatic. Multiple channels emerge in the transport as a positive magnetoresistance (MR) and a simultaneously non-linear Hall effect  ($\mu B\sim 1$, where $\mu$ is the mobility) and are described within two or three liquid models~\cite{Qu, Ren2010, Taskin}.
	In  quantum transport the  Dirac TSS must lead to the semi-integer quantum Hall effect, however, its observation is tricky because 3D TIs have two surfaces~\cite{Xu2014}. In view of quantum transport, TSS also 
	lead to positive magnetoresistance with a cusp-like minimum at $B = 0$ due to  a weak-antilocalization (WAL) suppression by the magnetic field. 
	An efficient number of conductivity channels  enters directly the WAL amplitude, i.e. the prefactor $\alpha_{WAL}$ value obtained from the Hikami-Larkin-Nagaoka (HLN) fit~\cite{Hik80}  of the low-field magnetoconductivity (MC). For a single channel $\alpha_{WAL}$ should be equal to -0.5, for two channels $\alpha_{WAL}=-1$, correspondingly.  When the bulk carriers contribute to MC, a nonzero interlayer coupling can arise, as a result,  experimentally, the  $\alpha_{WAL}$ often lies in the interval 
	between -0.5 and -1~\cite{Brahlek, Abad,  Assaf, Kuntsevich2016, Pandey2022}. 

 Apparently
  unphysical elevated values of $\alpha_{WAL}>1$ were   observed in a number of papers~\cite{Assaf,Wang15, Lang, Gornyi, Chen2011, Checkl, Shrestha}. For example, Ref.~\cite{Wang15} reports~$\alpha_{WAL}\sim$6. In several studies, these large $\alpha$ values were attributed to the classical linear~\cite{Parish, singh} or quadratic effects~\cite{singh22, Assaf}. 
 Nevertheless, in some papers~\cite{Wang15, Assaf}, large values of $\alpha_{WAL}$ persist even with these effects taken into account. 
 
 In this work we experimentally focus on the Hall effect and MC studies in the low-field domain in Bi$_{2-x}$Sb$_x$Te$_{3-y}$Se$_y$ (BSTS)  films, where WAL is observed. Quaternary compound BSTS were used to decrease the bulk concentration~\cite{Taskin, Xu2014, Lee2012, Xia2013, Tu2017}, which is high due to the doping by intrinsic point defects in binary 3D TI tetradymites,
such as Bi$_2$Se$_3$~\cite{xia} and Bi$_2$Te$_3$~\cite{Qu}. We reveal an elevated amplitude of the WAL and show that this  enhancement cannot be attributed to the electron-electron interaction or an admixture of linear or quadratic MR, as  was previously conjectured~\cite{Wang15, singh}.

Naive understanding and the theory of weak localization~\cite{Fukuyama, Altshuler} suggests that, in the low-field limit, the Hall resistance should be linear in the magnetic field. However, we observe the opposite: Hall coefficient  may vary by as much as 10-20$\%$.

 We note that Hall  nonlinearity has been previously observed in single crystals and thin films of 3D TIs and is always described in terms of multi-liquid model, i.e. parallel connection of several conductive channels with various densities $n_i$ and field-independent mobilities $\mu_i$ ~\cite{Qu, Ren2010, Taskin}. Within this model Hall effect deviates from the linear-in-field behavior in a magnetic field $>1/\mu_{max}$, where $\mu_{max}$ is the highest mobility among the components.  Since the maximal mobility ever reached in thin films of bismuth chalcogenides is about 1m$^2$/Vs $=1$ T$^{-1}$, one would not expect strong Hall nonlinearity below 1 T, that is the classically weak magnetic field for our films ($\mu B\ll 1$).

Small Hall effect nonlinearity in a classically low magnetic field was observed in various material systems  including 2D gases in GaAs~\cite{Minkov}, Si~\cite{Kuntsevich2013} and LAO/STO~\cite{Joshua}, doped semiconductors~\cite{Newson,Zhang1992}, indium oxide~\cite{Tousson}.  In all cases the effect depends strongly on temperature and the Hall feature sharpens in the low temperature limit. Several theoretical mechanisms  were suggested, including second order corrections~\cite{Minkov}, memory effects~\cite{Dmitriev2008}, superconducting fluctuations~\cite{Michaeli} and  combination of nonuniformity and quantum interference ~\cite{Kuntsevich2020}.  These theories (besides Ref.~\cite{Dmitriev2008}, which is not related to our case) also explain temperature dependent Hall nonlinearity.

We believe  therefore that all these mechanisms are irrelevant in our case of thin 3D TIs films. A large effect value, its elevated value in the lowest-density films and weak temperature dependence force us to suggest that the effect is related to the (i) low-field transport current redistribution  due to  conductivity  drop in one channel and (ii) emergence of  this drop  from the the magnetic field effect on the electron  spectrum.

We suggest that the anomalous WAL amplitude and low-field Hall feature are interrelated and originate from the suppression of the surface states contribution to the conductivity by the magnetic field. Elevated Hall conductivity of the surface states may be either due to low concentration or due to an anomalous Hall effect.  In our model the Fermi level should be close to the Dirac point of the surface states. The magnetic field opens the Zeeman gap and suppresses the conductivity of the surface states and their contribution to the transport. This model qualitatively agrees to  the data and, thus, is a candidate for explanation of the observed features.

Nevertheless, the physical reason for the Hall nonlinearity and large WAL amplitude remains not firmly established. Our results, thus, draw the  attention of theory and experiment to these phenomena.

\section{Samples and experiment}

Bi$_y$Sb$_{2-y}$Te$_{3-x}$Se$_x$ (BSTS) films were grown by molecular beam epitaxy on a Si(111) substrate. The details of growth and structural characterization
were described in Ref.~\cite{Ste_cr}.  
Along with the compound BiSbTeSe$_2$ source, the elemental Bi and elemental Te sources were used. After obtaining the Si(111)-7x7 surface reconstruction, the sample was cooled to the growth temperature and kept at this temperature for one hour to reduce the surface temperature gradient on the sample. To passivate the silicon dangling  bonds, which reduce the adatom diffusion length  and lead to the quality degradation of the A$^V$B$^{VI}$ compound growing layer, the tellurium source was opened for  3 - 5 minutes. The saturation of dangling bonds with tellurium atoms was evidenced by a decrease in the intensity of the Si(111)-7 x 7 surface structure. Then, a 2-4 nm thick Bi$_2$Te$_3$ buffer layer was grown with the appearance of a continuous film being controlled by the emergence of a streaky RHEED pattern.
The substrate temperature T$_{sub}$, temperature of sources T$_{Te}$, T$_{Bi}$ and T$_{BSTS}$ and growth rate  were changed (T$_{sub}$ - between 345 and 370$^\circ$C, T$_{BSTS}$ was 230-240$^\circ$C, growth rate $\nu$ was changed in the range 0.07-0.5 nm/min) to find the optimal growth parameters. 
According to the atomic force microscopy data, the total   Bi$_y$Sb$_{2-y}$Te$_{3-x}$Se$_x$ film thickness is varied between 12 and 60 nm.

The structures for the transport measurements were prepared  by both the shadow mask technique, where a standard six- terminal Hall bar mask   was placed on   top of a substrate, 
and  by optical photolithography with  HNO$_3$:CH$_3$COOH:H$_2$O$_2$:H$_2$O etchant being used to form a  mesa-structure on BSTS films. In the first case  indium contacts were sputtered atop; in the second one Ti/Au contacts were defined with the lift-off technology. 
\begin{figure}
	\centering
	\includegraphics[width=3.0in]{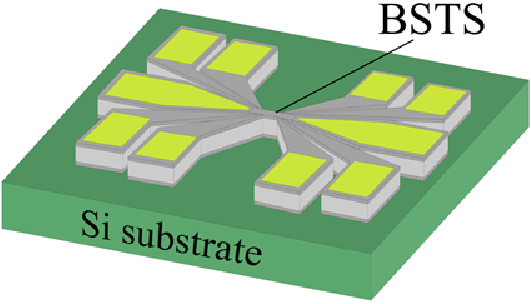}
	\caption{\label{structure} Scheme of the structure with  lithographically produced contacts. The gray color corresponds to the surfaces of BSTS, light gray corresponds to the BSTS bulk. }
	\end{figure}

 In Fig.~\ref{structure} is a lithographically formed structure. Both In and Ti/Au contacts show the same transport characteristics.    The magnetoresistance and temperature dependences of conductivity  were measured in a helium Dewar in magnetic fields up to 4 T and also in the Oxford dry cryomagnetic system up to 12 T and temperature down to 1.64 K. The magnetic field was swept from positive to negative values and all magnetoresistance (Hall resistance) curves were symmetrized (anti symmetrized) to account for the misalignment of the potential contacts. 

\section{Experimental results}
\subsection{Observation of enhanced WAL prefactor}
The temperature  dependencies of the resistivity $R(T)$ for five samples  under study (summarized in Table~\ref{tabl}) measured up to the liquid nitrogen temperature are shown in Fig.~\ref{RvsT}. (Typical  $R(T)$ dependence measured to higher temperature is given in Supplemental materials~\cite{sup} for the  sample 36, Figure S1 ). 
For all samples the resistance is weakly changed with temperature, not more than 5$\%$, 
that indicates low carrier mobility and is rather general for 3D TI thin films. The nonmonotonic character of temperature dependence  is  apparently due to the competition of several transport mechanisms: quantum corrections to conductivity, i.e. electron-electron correction and weak antilocalization, having different functional dependencies from bulk of the film and surface states, Fermi level drift, phonon scattering.
The low temperature part of the $R(T)$ curve and the observed resistivity upturn at a temperature below  10-15 K are usually explained in a 3D TI by a collective contribution of the WAL, weak localization (WL) and electron-electron interaction (EEI) corrections to the conductivity from the surface and bulk  channels ~\cite{sas21, wang2011, Kuntsevich2016}. 

\begin{figure}
	\centering
	\includegraphics[width=3.0in]{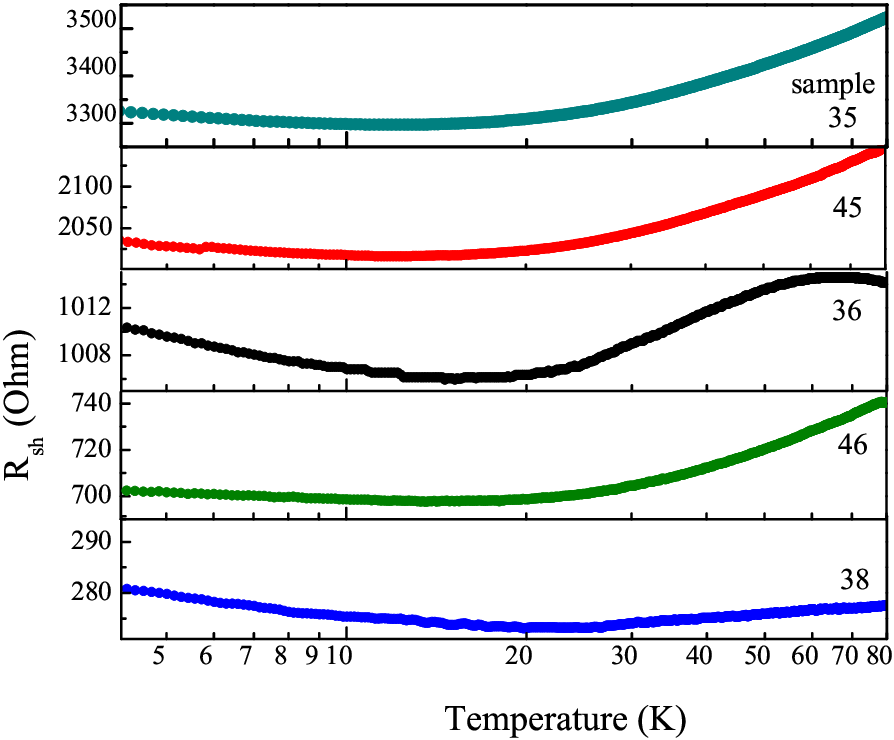}
	\caption{\label{RvsT} 
		Temperature dependence of resistance for different BSTS films.}
\end{figure}
\begin{figure}
	\centering
	\includegraphics[width=3.0in]{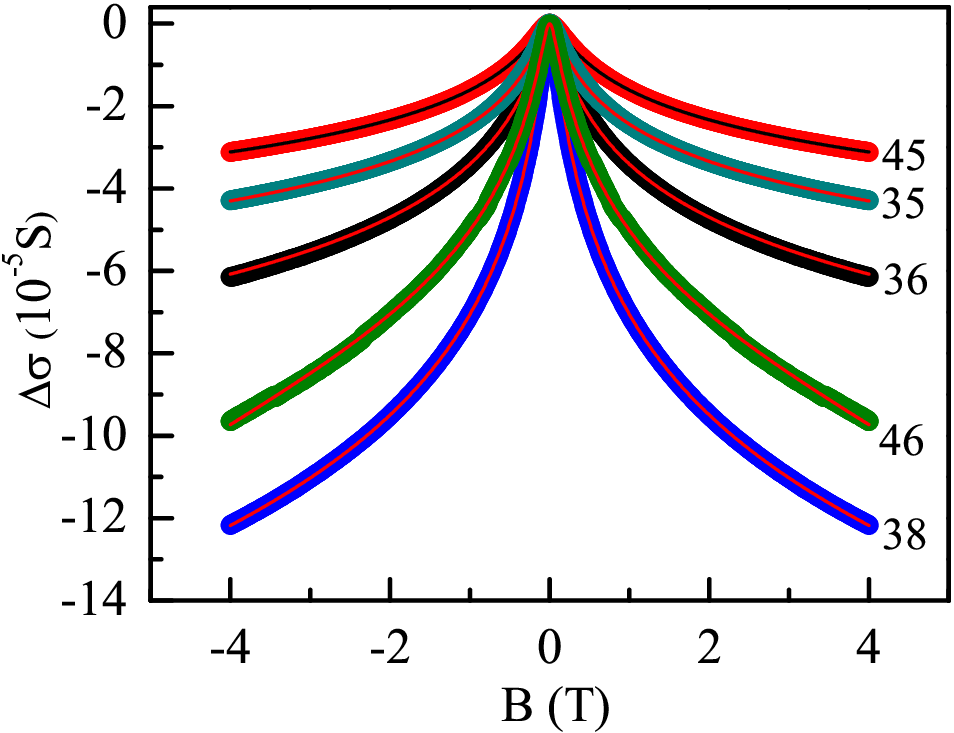}
	\caption{\label{MC4samples} 
		Magnetoconductance $\Delta\sigma(B)=\sigma(B)-\sigma(0)$ of the samples under study in the magnetic field perpendicular to the sample plane, T=4.2 K. Lines -- approximation with the HLN formula.}
\end{figure}

The low-temperature magnetoconductance (MC) for all  films is similar and negative with a cusp-like maximum at $B = 0$ (Fig.~\ref{MC4samples}). Such behavior in TIs is usually associated with the suppression of WAL  correction to the conductivity of the surface states. 

We fit
the observed MC curves and $\Delta\sigma(B)=\sigma(B)-\sigma(0)$ in the conventional way~\cite{Hik80} using the  Hikami-Larkin-Nagaoka type (HLN) formula:

\begin{equation} \label{HLN}
\Delta\sigma(B)=\alpha\frac{e^2}{\pi h}\left[\Psi\left(\frac{B_0}{B}+\frac{1}{2}\right)-\ln\left(\frac{B_0}{B}\right)\right]
\end{equation}   
Here $B_0 = \hbar/4e(L_\varphi)^2$, $L_\varphi$ is the phase coherence length,
$\alpha$ is a constant  that should be equal to $-1/2$ for WAL with one conducting channel and
$\Psi(x)$ is the digamma function.

An explicit comparison of the experimental
results to the HLN theory is shown by the solid lines in Fig.~\ref{MC4samples}.
The experimental data for perpendicular field MC  are apparently well described by the WAL corrections, though  the values of $\alpha_{WAL}$ are unphysically large (see Table ~\ref{tabl}). Our data are not surprising since the elevated values of $\alpha_{WAL}$ were previously  observed in a number of papers \cblue{~\cite{Assaf,Wang15, Lang, Gornyi, Chen2011, Checkl, Shrestha}}.

\begin{figure}
	\centering
	\includegraphics[width=3.0in]{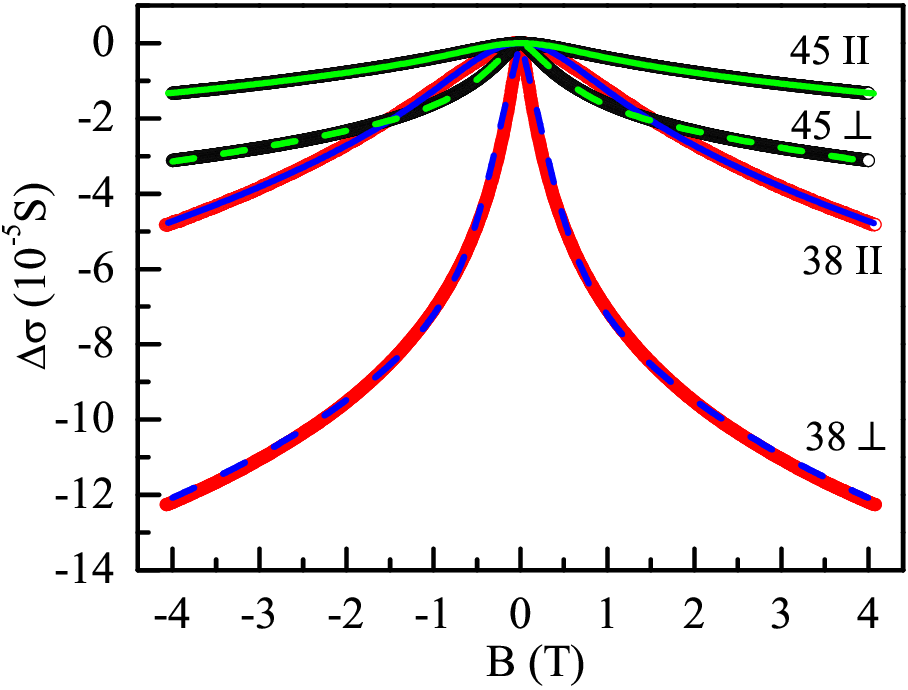}
	\caption{\label{fig8} 
		Magnetoconductance $\Delta\sigma(B)=\sigma(B)-\sigma(0)$ for  samples 45 and 38 in the magnetic field perpendicular and parallel to the sample plane, $T=4.2$ K. Solid lines correspond to the EEI magnetoresistance, dashed lines correspond to the the calculations with Eqs.~(\ref{ee0}-\ref{R2liq}) that include EEI, WAL and the gap opening mechanism.} 
\end{figure}
\begin{table}[ht]
	\centering
		\caption{\label{tabl}Sample characteristics: film thickness $d$, mean free path $l$,   parameters $\alpha_{WAL}$ , $L_\phi$ obtained from the fitting of the conductance data by the Eq.~\ref{HLN}, Hall concentration n$_H$ and mobility $\mu$ for averaged in 2-4 T magnetic field   Hall coefficient, measured at 4.2 K.  }

\begin{tabular}{|c|c|c|c|c|c|c|c|}
\hline	
		Sam- & $\alpha_{WAL}$ & $L_\varphi$ & $k_F\times l$ & $l$  & $d$&$ n_H$& $\mu$ 
		\\
			ple& & nm & &   nm  & nm & cm$^{-2}$& cm$^2$/Vs\\
		\hline
\verb"45" &-1.01 &58.2&7.8&10&12&9.4e12&200\\
	\verb"46"& -2.65 & 69.3 & 33 & 39 & 18 &1.1e13 & 709\\
		\verb"35"& -1.19 & 72.2 & 9 & 9.7 & 14 &1.3e13 & 159\\
\verb"36"&-1.69&72.4&20.4&13.2&60&3.8e13&130\\
	\verb"38"&-3.11&81.3&34&16.5&35&6.9e13&121\\

	\hline
\end{tabular}
\end{table}

	 It is often conjectured that additional quadratic contribution to conductivity or linear magnetoresistance (LMR) due to macroscopic inhomogeneity of the film \cite{Parish} may modify the prefactor~\cite{Wang15, Assaf, singh22, Stephen}. 
		 We carefully checked numerically (see Appendix) that it is not the case for our dataset:  $\alpha_{WAL}$ value modifies slightly if the magnetoresistance is fitted by a sum of Eq.\ref{HLN} and a parabolic or linear term. The elevated prefactor in Eq.\ref{HLN} thus suggests for an additional MC mechanism acting in parallel with WAL and leading to a sharp positive magnetoresistance.

\subsection{High-field magnetoresistance. The role of electron-electron interaction correction}	 \label{sec:MR} 	   	  	
	
	The high-field part of MC observed for  all our samples as well as in some other studies~\cite{singh, X.Zhang, Stephen} is not typical of WAL, because the magnetic fields, in which the WAL effect must vanish, $B_{tr}\sim hc/el^2$, where $l$ is a mean free path, should be less than 1 T.  
 
 Both the WL/WAL and macroscopic inhomogeneity mechanisms of magnetoresistance \cite{Parish} are orbital effects and should be suppressed when the magnetic field is parallel to the TI film.   To elucidate the nature of the high-field smooth $R(B)$ dependence with respect to magnetic field,  the sample rotation was carried out. As shown in Fig.~\ref{fig8} for the samples 38 and 45 that have vastly  different  $\Delta\sigma$ and resistance values, the high-field part of magnetoresistance  is weakly sensitive to the field direction. It means that the EEI corrections to the conductivity  have to be taken into account. 
The EEI in the magnetic field must be affected by Zeeman splitting~\cite{AltAronov}.  This effect should be large enough due to the large $g$-factor, typically from 10 to 60 in tetradymite-based TIs~\cite{wolos, Koh, Anali, Taskin,Liu2010}. 

Negative EEI magnetoconductance
$\Delta \sigma_{EEI}(B)$ appears when the spin splitting by an external field exceeds the temperature. In high magnetic fields ($g\mu_BB/k_BT \gg $1), 
$\Delta \sigma_{EEI}$ has a logarithmic-in-$B$ dependence for 2D films and $\propto \sqrt{B}$ in the 3D case~\cite{EEIMR}.  In arbitrary fields in 3D it is described with the expression:
\begin{equation}\label{sigB}
\sigma_b(B) = \sigma_b^{(0)} - A_{EEI} \times g_3\left(\frac{\mu_b g B}{T}\right).
\end{equation}
Here $g_3(x)\propto \sqrt{x}$ for $x\gg 1$ is the dimensionless function.  $A_{EEI}$ is the amplitude of EEI magnetoresistance that depends on the diffusion coefficient in the bulk, the screening of Coulomb potential and the film width. The $g$-factor may depend on the  magnetic field orientation (parallel or perpendicular to the film) but $A_{EEI}$ cannot.

The 3D EEI mechanism reasonably agrees to the high field part of magnetoresistance data in parallel magnetic field (solid lines in Fig.~\ref{fig8}). The fitting parameters for sample 38 are $g_{\|}=29.4$, $A_{EEI}=1.56\cdot 10^{-5}\,{\rm S}$ and for sample 45: $g_{\|}=35.6$, $A_{EEI}=0.38\cdot 10^{-5}\,{\rm S}$. Here $g_{\|}$ is the bulk $g$-factor in the parallel field. These values are in agreement with the typical $g$-factors measured  in tetradymite-based TIs~\cite{wolos, Koh, Anali, Taskin,Liu2010}. 
We, thus, presume that the magnetoresistance in high perpendicular fields is driven also by EEI, and this correction originates from the bulk of the film.

  In a perpendicular magnetic field the orbital mechanisms of magnetoresistance should come into play.  We showed that the magnetoresistance in very small perpendicular fields is described by WAL with an unphysically large prefactor $\alpha_{WAL}$. Adding up EEI does not help to make the WAL amplitude reasonable.
 It means that, besides WAL and EEI, there is an additional mechanism of magnetoconductance,  specific for our films.  We believe that the key to the understanding of this mechanism is the Hall effect measurements described in the next section.

 \subsection{Low-field Hall nonlinearity}

\begin{figure}
	\centering
	\includegraphics[width=3.0in]{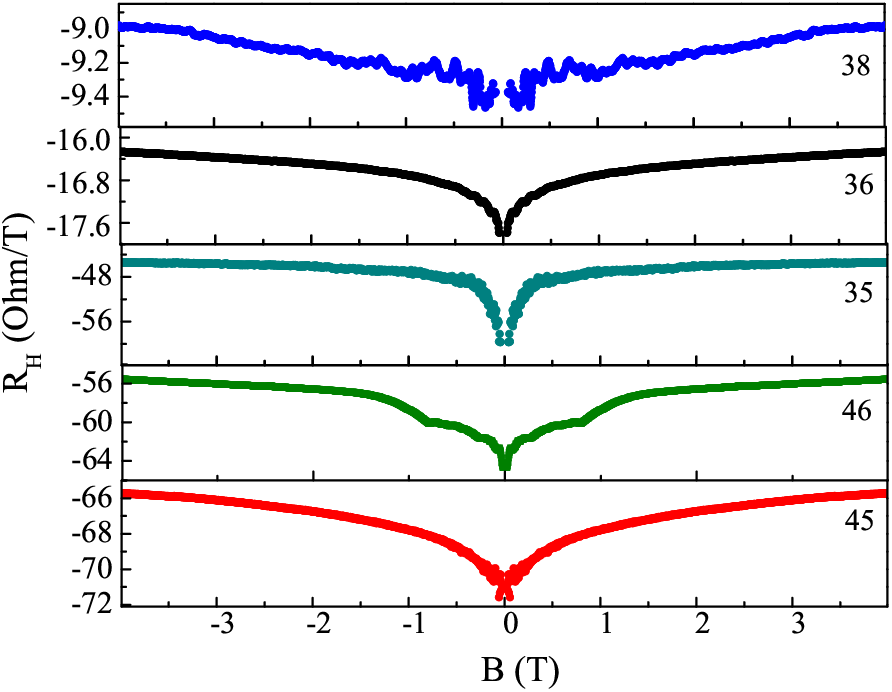}
	\caption{\label{fig6} Hall coefficients for the samples under study measured at 4.2 K. }
\end{figure}

 We measured the Hall resistance $R_{xy}(B)$ and focused on the Hall coefficient $R_H(B)\equiv R_{xy}(B)/B$. Fig.~\ref{fig6} demonstrates the  $R_H$(B) dependence for the samples under study measured at 4.2 K. One can see that the $R_H$ dependence is non-linear and 
 strong in classically weak magnetic fields $\lesssim$~1~T.

 \begin{figure}
	\includegraphics[width=3.0in]{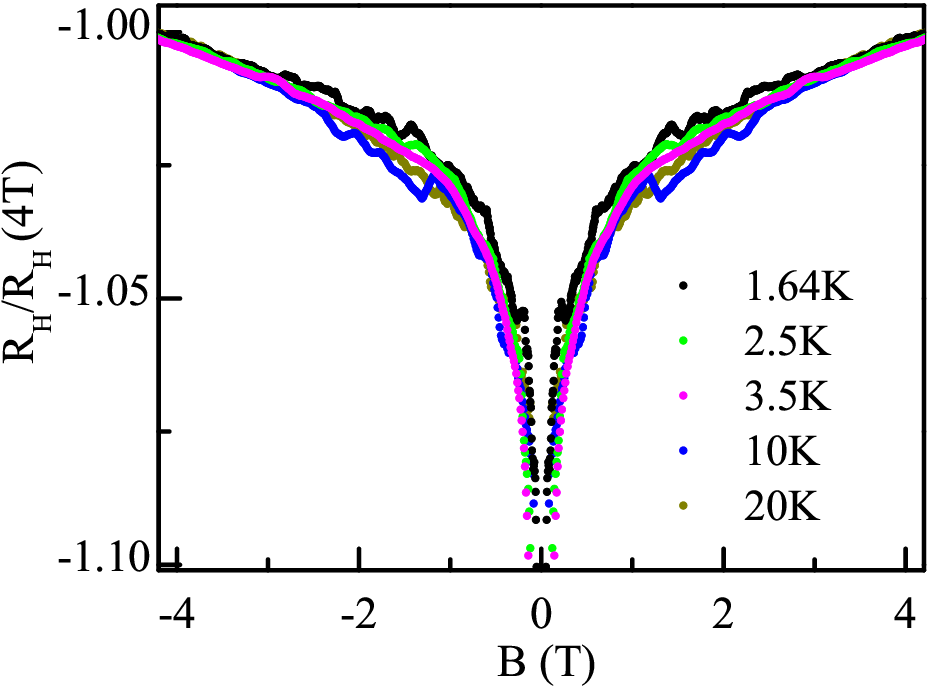}
	\caption{\label{fig9}  Temperature dependence of Hall coefficients $R_H$ for  sample 36 normalized to the high field $R_H$(4T) value.   }
\end{figure}

{The scale of the effect differs from sample to sample. 
In the high carrier density of sample 38 there is  almost no visible nonlinearity. However, the effect becomes essential in the lower density samples,  
where the TSS do contribute to the conductivity. 

 In higher magnetic fields  
the Hall coefficient smoothly continues to decrease, also in  sample 38. This might be due to classical effects, such as two-liquid redistribution and field dependence of mobility. In our paper, we focus on the low-field feature.

Important is that the effect does not demonstrate a visible temperature dependence in the low temperature domain $T\lesssim~$20~K (see Fig.~\ref{fig9}). It means that we may rule out most of the mechanisms so far suggested to explain the low-field Hall non-linearity: WL-related current redistribution in inhomogeneous systems\cite{Kuntsevich2020}, second order correction from WL/WAL and EEI\cite{Minkov} and superconducting fluctuations\cite{Michaeli}. In uniform systems, WL and WAL do not affect the Hall slope~\cite{Fukuyama}. 

The absence of the temperature dependence along with the large value of the effect assume that its nature is related to the current flow redistribution: in elevated magnetic fields the transport current bypasses the regions, that  strongly contribute to the Hall effect at a small magnetic field. } 
   It was natural to assume that  low-field magnetoresistance, much exceeding the theoretical prediction, arises in only one component (e.g. topological surface states). A drop of its  conductivity with the field, according to the two-liquid model, leads to a decrease of the fraction of the corresponding transport current. Since the TSS had a low density, the overall Hall coefficient should decrease.

 \section{Model}
\label{sec:model}

\begin{figure}

\includegraphics[width=3.0in]{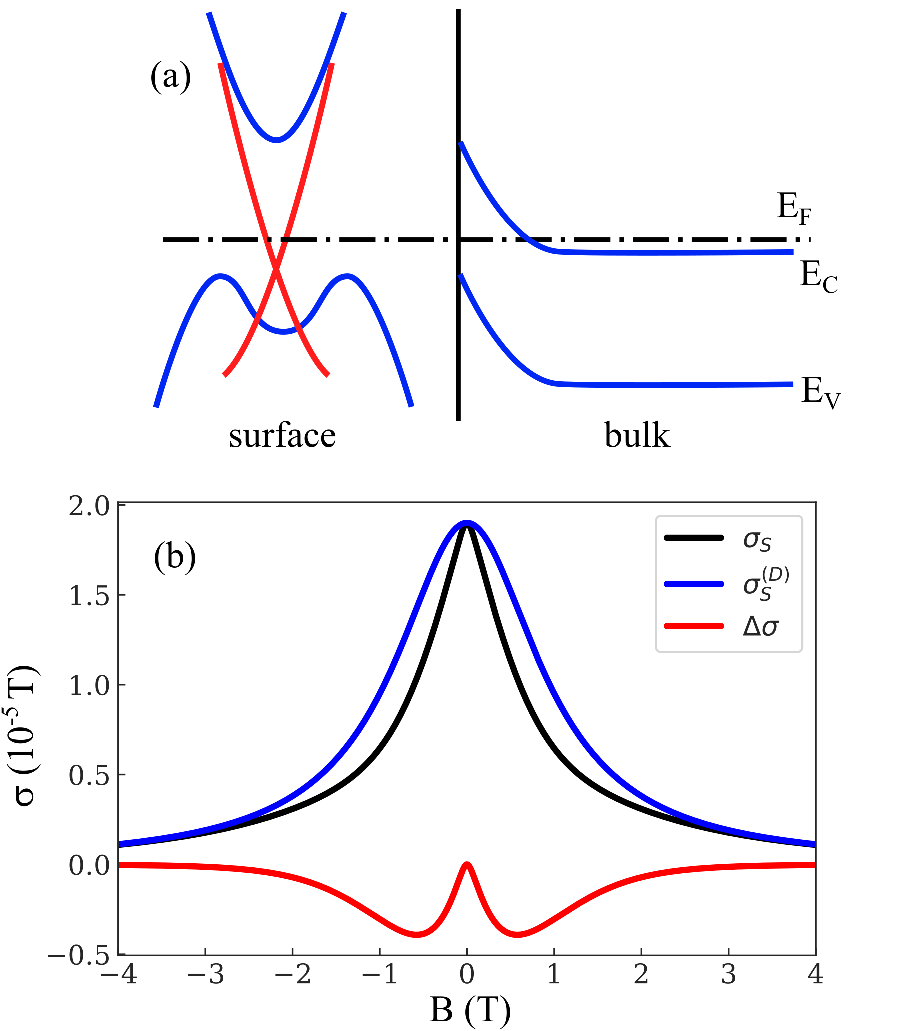}
\caption{\label{figModel} 
Illustration of the model. (a) The energy spectrum of the bulk and surface states at zero magnetic field. (b) Drude surface conductivity $\sigma_s^{(D)}$, WAL correction $\Delta\sigma$ and the resulting surface conductivity $\sigma_s$ for sample 45.} 
\end{figure}

In this section, we show that both the elevated value of $\alpha$ and Hall non linearity can be explained within  a model based on the following assumptions. (1) The chemical potential at the surface is close to the Dirac point. The surface electron concentration is small,  the surface conductivity is of the order of the conductivity quantum \cite{Culcer, Akzyanov}. The Hall conductivity of the surface in the magnetic field is high either within the Drude model or within the anomalous Hall mechanism\cite{anoHall}. (2) The chemical potential is inside the conductive
band due to the strong band bending shown in Fig.~\ref{figModel}(a). The surface states are separated from the bulk with the tunnel barrier, therefore the two kinds of states act as independent channels.
(3) The second surface (Si-BSTS) does not contribute to the magnetoresistance and Hall effect (see Discussion).

The surface states are described with TI Hamiltonian
\begin{equation}\label{Ham}
\hat{H}= v_0 (\sigma_x p_y-\sigma_y p_x) + g_s\mu_B \bm{\sigma} {\bf B}
\end{equation}
Here $v_0$ is the Fermi velocity at the Dirac point, $p_{x,y}$ is the surface electron momentum, $g_s$ is the surface electron $g$-factor that can be different from the $g$-factor in the bulk. 

In a zero magnetic field, the surface states form a Dirac cone. We presume that the Fermi energy is shifted from the Dirac point by the small energy $\varepsilon_F^{(0)}\sim 
{10^{-2}}\, {\rm eV}$ allowing the suppression of surface conductivity with a relatively low perpendicular magnetic field $B_z$ that leads to the gap opening. The dispersion with finite $B_z$ is as follows
\begin{equation}
\varepsilon(p) = \pm \sqrt{v_0^2 p^2 + \varepsilon_Z^2}.
\end{equation}
Here $\varepsilon_Z = \mu_b g_s B_z /2$ is Zeeman energy in the perpendicular field. Note that the parallel field only shifts the Dirac point but does not open the gap making the effect anisotropic. 

The gap increases the effective mass and pushes the electrons towards the bulk states. This process is compensated by the apparent electric surface charge that results in the increase of the surface Fermi energy. When this compensation
is strong, the 2D electron density is almost conserved and the Fermi energy is modified in the magnetic field \cite{nconst}
\begin{equation}\label{ee0}
\frac{ \varepsilon_F}{\varepsilon_F^{(0)}} =\sqrt{1 + (\varepsilon_Z/\varepsilon_F^{(0)})^2}. 
\end{equation}
Here we presume that the Zeeman energy is larger than the temperature.

The exact expression for the conductivity due to the emergence of electron mass depends on the particular scattering mechanisms. We consider the scattering by neutral impurities, which leads to the transport time $\tau_{tr}\propto \varepsilon_F^{-1}$. The Fermi mass $m\propto \varepsilon_F$ results in the Drude conductivity dependence on the Fermi energy  
\begin{equation} \label{sigmaD}
\sigma_s^{(D)}(B) = \sigma_s^{(D)}(0) \frac{\varepsilon_F^2(0)}{\varepsilon_F^2(B)} 
\end{equation}
We presume that the decrease of the Drude conductivity is responsible for the magnetoresistance in  intermediate perpendicular fields $B_z \sim 0.5\, {\rm T}$. 

 The spectrum modification also changes the WAL.
In the low field the WAL correction to the magnetoconductivity is significant due to sharp $B$-dependence $\sigma_s(B) = \sigma_s^{(D)}(B) + \Delta\sigma(B)$. $\Delta \sigma(B)$ is described by Eq.~(\ref{HLN}) with a single channel and $\alpha=-1/2$, 
which we do not treat as a fitting parameter. 

The phase coherence length $L_{\varphi}$ for the surface states is equal to $L_{\varphi} = \sqrt{D\tau_{\varphi}}$. Here $D\sim v_F^2 \tau_{tr}$ is the diffusion coefficient,  $\tau_{\varphi}$ is the phase coherence time and $v_F$ is Fermi velocity (that can be different from $v_0$ due to the gap opening). The EEI leads to $\tau_{\varphi} \approx \hbar^2 g(\varepsilon_F) D/T$~\cite{AltAronov}, where $g(\varepsilon_F)$  is the density of states at the Fermi level.
Eq.~(\ref{HLN}) is valid up to the transport field $B_{tr} \sim \hbar^2/4eD\tau_{tr}$~\cite{Golub}.

Both $D$ and $g(\varepsilon_F)$ are affected by the gap opening and increase of the apparent mass of surface electrons. It leads to the renormalization 
of $B_0$ 
in Eq.~(\ref{HLN})
\begin{equation}
B_0(B) = B_0(0)\frac{\varepsilon_F(B)}{\varepsilon_F(0)}
\left(
1 - \frac{\varepsilon_Z^2}{\varepsilon_F^2(B)}
\right)^{-2}
\end{equation}

Drude surface conductivity $\sigma_s^{(D)}(B)$, the WAL correction $\Delta\sigma$  and the total surface conductivity $\sigma_s$ are shown in Fig.~\ref{figModel}(b). 

The sheet resistivity and the Hall coefficient are calculated within  the two liquid model (bulk with EEI and surface with WAL). Note that such model is valid when the bulk-surface scattering length is larger than the scattering lengthscales of the surface.

\begin{equation}\label{R2liq}
\sigma = \sigma_b + \sigma_s, \quad 
R_H = \frac{R_s \sigma_s^2 + R_v\sigma_v^2}{(\sigma_s + \sigma_v)^2}.
\end{equation}
It shows that the sample conductivity $\sigma$ is just the sum of $\sigma_b$ and $\sigma_s$. The sample Hall coefficient $R_H$ is some function of surface Hall coefficient $R_s$ and bulk Hall coefficient $R_b$. It also depends on conductivities $\sigma_b$ and $\sigma_s$.

In Fig.~\ref{fig8} is the quantitative agreement between the perpendicular field magnetoconductance and the model including the  EEI, WAL and gap mechanisms (dashed lines). To achieve it, we first compared the high-field part of magnetoconductance with EEI (Eq.~(\ref{sigB})), presuming that other mechanisms are saturated at $B\gtrsim 3\,{\rm T}$. In this analysis the amplitude $A_{EEI}$ was considered to be known from the parallel field magnetoconductance. We presume that the $g$-factor in perpendicular field $g_{\perp}$  differs from $g_{\|}$. The best agreement between the magnetoconductance measured in the high field and Eq.~(\ref{sigB}) is reached for  $g_\perp = 39.4$ in  sample 38 and $g_\perp = 56.9$ in  sample 45. Then we compare the magnetoconductance in arbitrary fields with the two-liquid  model. $\sigma_s(0)$, $B_0$, and $\varepsilon_F^{(0)}$ were considered as the fitting parameters. The dashed curves in Fig.~\ref{fig8} correspond to $\sigma_s(0)=7.4\cdot10^{-5}\,{\rm S}$, $B_0 = 0.006 \,{\rm T}$, $\varepsilon_F^{(0)}/\mu_b g_s = 0.6 \,{\rm T}$ for the sample 38 and to $\sigma_s(0)=1.75\cdot10^{-5}\,{\rm S}$, $B_0 = 0.035 \,{\rm T}$, $\varepsilon_F^{(0)}/\mu_b g_s = 0.9 \,{\rm T}$ for the sample 45. 
Note that, in both samples the surface conductance is comparable with $\sim e^2/h$. 

 The agreement between the experiment and Eq.~(\ref{R2liq}) requires a rather small difference $\varepsilon_F^{(0)}$ between the Fermi level at the surface and the Dirac point. It should be comparable to the Zeeman energy in the field $\sim$1 T.  With respect to a large g-factor $\sim 100$,  it corresponds to $\varepsilon_F^{(0)}\sim 0.01 \, {\rm eV}$. Otherwise, if Fermi level at the surface would lie far from the Dirac point, Eq.~(\ref{sigmaD}) results only in  the correction to the conductivity that is  quadratic over the magnetic field. It would be hardly distinguishable from the other mechanisms of quadratic magnetoconductance.

\begin{figure}
	\centering
	\includegraphics[width=3.0in]{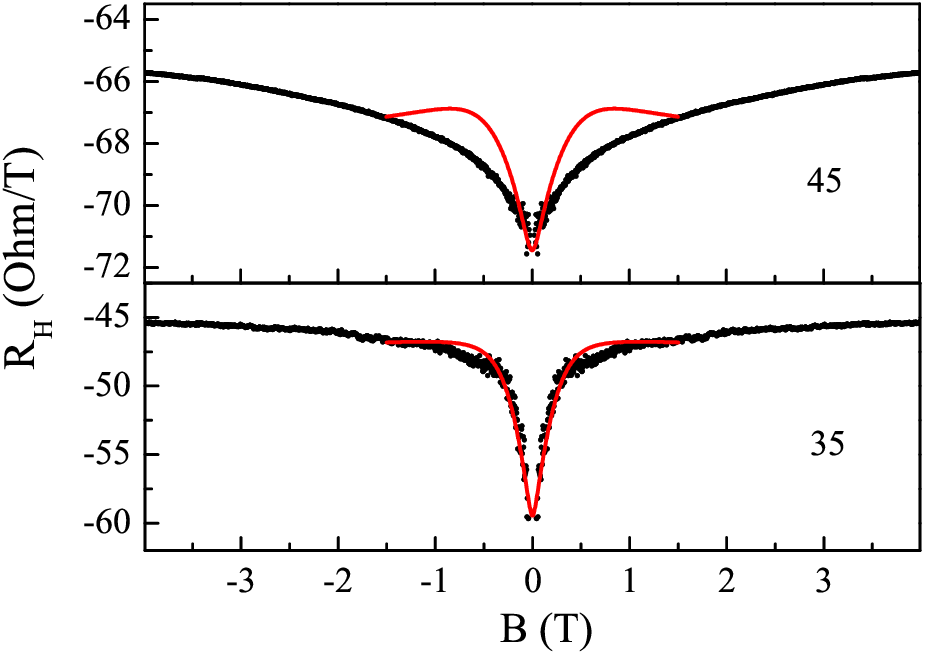}
	\caption{\label{fig:fitR} 
		Hall coefficient measured at 4.2 K for samples 35 and 45, compared with Eq.~(\ref{R2liq}).\ }
\end{figure}

Fig.~\ref{fig:fitR} compares the Hall coefficient for samples 45 and 35  in the low magnetic field with the prediction of the model. From the magnetoconductance analysis we get the following values for  sample 35: $g_\perp = 48.14$, $\sigma_s(0)=2.05\cdot10^{-5}\,{\rm S}$, $B_0 = 0.02 \,{\rm T}$, $\varepsilon_F^{(0)}/\mu_b g_s = 0.5 \,{\rm T}$. The values $R_s$ = -3220 and $R_b = -68.3$ are used for fitting the Hall coefficient for  sample 45 and $R_s$ = -5100 and $R_b = -47.2$ for  sample 35. 

{Note that the agreement between the experiment and the model requires $R_s$ to be much larger than $R_b$. It corresponds to the assumption of low surface states Fermi energy. The classical Hall coefficient is large for the small carrier density, while the anomalous Hall coefficient in TIs\cite{anoHall} is large for small Fermi energy $\varepsilon_F(0)$. Therefore, both mechanisms should lead to the anomalously strong contribution of surface transport to the Hall effect. }

 The  qualitative agreement between experiment and model is shown in the figure.

\section{Discussion.}

{In our paper we experimentally revealed a low-field Hall effect nonlinearity in 3D TI thin films. Our data differ from the previously reported observations of such nonlinearity in semiconducting 2D systems by a relatively large amplitude and weak temperature dependence. These signatures  motivate us to believe that we deal not with coherent effects, but rather with the transport current  redistribution due to the modification of the electron spectrum  in the magnetic field. Namely, the system has several conductive components acting in parallel. The mobility of some of them (presumably Dirac surface states) drops with the magnetic field. At zero field these states possessed a rather low density and high mobility and led to an elevated Hall constant. The magnetic field, therefore, promotes a drop of the Hall constant.

We note that, besides the bulk, there are two surfaces in 3D TIs thin films- top  and bottom ones.  These surfaces are not equivalent that further complicates the problem. As a rule, the bottom surface is much more disordered due to the lattice mismatch and scattering~\cite{chen10}. 
An indirect indicator of different surfaces is the observation of WAL with prefactor $\alpha=0.5$ instead of $\alpha=1$ in  most of 3D TI thin film samples \cite{GraciaAbad, Chen2011}.   In some cases of well adjusted 3D TI  substrates, e.g. GaSe~\cite{He_sciRep}, or graphene \cite{Ste_gr}, both surfaces contribute equally.  We do not  definitely know how many surfaces are effectively involved in our structures, however, we checked that, for a qualitative comparison the effective number of surface channels (1 or 2) is not essential.

Due to classically weak magnetic fields we simplified the problem, neglected  Landau quantization effects and considered the theoretical effect of the Zeeman field-induced gap opening for the Dirac states of 3D TI thin films. For  realistic $g$-factor values   ($\sim 50$) this mechanism, along with WAL and EEI corrections, qualitatively explains much of experimental data:
\begin{enumerate}
    \item The elevated value of the WAL prefactor $\alpha_{WAL}$ obtained from the Hikami-Larkin-Nagaoka fit of low field magnetoresistance. This value is high because the current redistribution contributes to positive magnetoresistance. 
    \item Almost temperature independent Hall nonlinearity driven by the spectrum modification and current redistribution.
    \item High-field monotonic magnetoresistance (almost the same in the  parallel and perpendicular fields) due to electron-electron interactions.

\end{enumerate}

 The suggested mechanism  involves a rough assumption of a specific surface band bending, so the Fermi level is close to the Dirac point. We can not prove that it is indeed the case experimentally and can not exclude that a more universal mechanism of the phenomenon will be found.

Note that the Hall non-linearity is not usually reported. We believe there are two reasons for this:  (i)  researchers do not pay enough attention to the low field limit for the Hall effect because they do not expect any nonlinearity and (ii) it is not  easy to obtain thin films with a rather low carrier density, Fermi level close to the Dirac point for the topological surface states and, at the same time, with some bulk conduction electrons. 
For example,  in Bi$_2$Te$_3$, the Dirac point is below the top of the valence band.  In Bi$_2$Se$_3$ the Dirac point is in the middle of the 0.3 eV band gap, too far from the bottom of the conduction band. BSTS is probably ideal, therefore, for such observations.

We should also discuss here previous observations of elevated WAL prefactor $\alpha_{WAL}>1$ in relation to our data.  
 Recent work~\cite{Gornyi} reports $\alpha_{WAL}\sim$ 1.4  in BSTS microflakes, for the Fermi level close to the charge neutrality point (CNP). Away from the CNP $\alpha_{WAL}$ approached the theoretical value $\sim$ -1. Similar behavior was observed in Ref.~\cite{Lang} where the authors studied gate dependence of the conductivity and MR for the  (Bi$_{0.57}$Sb$_{0.43}$)$_2$Te$_3$ TI films around Dirac point. 
 These observations are in line with the mechanism, suggested in our paper.
 
 The elevated values of $\alpha_{WAL}$ were also reported in~ Refs.\cite{singh22, shekhar, Sahu, Assaf}. 
Usually the observed values of $\alpha_{WAL}$ and its 
temperature dependence are discussed in terms of several conductivity channels and their interaction. However, the most expected result of channel interaction is their merging that would lead to the decrease of $\alpha_{WAL}$~\cite{MultiCh} and $|\alpha_{WAL}| > 1$ is hard to achieve with only two channels with WAL.

   In Refs.~\cite{Sahu, Assaf} large values $|\alpha_{WAL}| > 1$ appear only at elevated temperatures, while in Refs. \cite{singh22, Lang} they are seen at low temperatures. In \cite{singh22} $\alpha_{WAL}(T)$ decreases with $T$. Our Eq~(\ref{sigmaD}) is derived in the limit $T \ll \varepsilon_F(0)$, at larger temperatures it should be suppressed. We do not think therefore that all the reports of elevated $\alpha_{WAL}$ could be explained by a single physical mechanism. It is possible that our theory is related to results of Refs.~\cite{singh22,Lang}, but not  Refs.~\cite{Sahu, Assaf}. We believe that the measurements of the Hall effect could be a useful tool to gain further insight. 

 The Hall non-linearity in Bi$_2$Te$_3$ in magnetic fields $B\sim 1\, {\rm T}$ was observed in Ref.~\cite{Qu} and attributed to the very high mobility of the surface states $\mu \approx 10^4 {\rm cm^2/Vs}$, which leads to $\mu B>1$ already in low fields. For our low mobility films this is definitely not the case. This assumption in our case would inevitably lead to very low Fermi energy and the importance of gap opening. 

We believe that our paper will help to draw attention to the low field domain and stimulate  theory development, and further experiments. In particular, the gap opening and current redistribution could be visualized by scanning probe techniques. Study of Hall effect in gate dependent measurements will also tell a lot about the effect, because such measurements allow  tuning the Fermi level at least for one of the surfaces.} \\

\section{Conclusion}
In this paper we report on the anomalously large weak antilocalization and nonlinear Hall effect in thin films of 
BSTS 3D topological insulators. The Hall feature emerges in classically weak magnetic fields and weakly depends on the temperature. The existing theories fail to explain the observed phenomena. 
We suggest a possible mechanism of magnetoconductivity, related to the Zeeman field induced gap opening at the Dirac point, for the surface states in the perpendicular magnetic field. This gap decreases the conductivity of the surface states, and, hence, their contribution to the Hall effect.
Our mechanism qualitatively explains the observed phenomena, although the validity of the assumptions is yet to be confirmed.

\section{Acknowledgements}
The authors are grateful to A.V. Nenashev for usefull discussion.
This work
was supported by the Russian Science Foundation  and Government of the Novosibirsk region (grant 22-22-20074).

\appendix

\section{Hall resistance}

\begin{figure}
	\centering
	\includegraphics[width=3.0in]{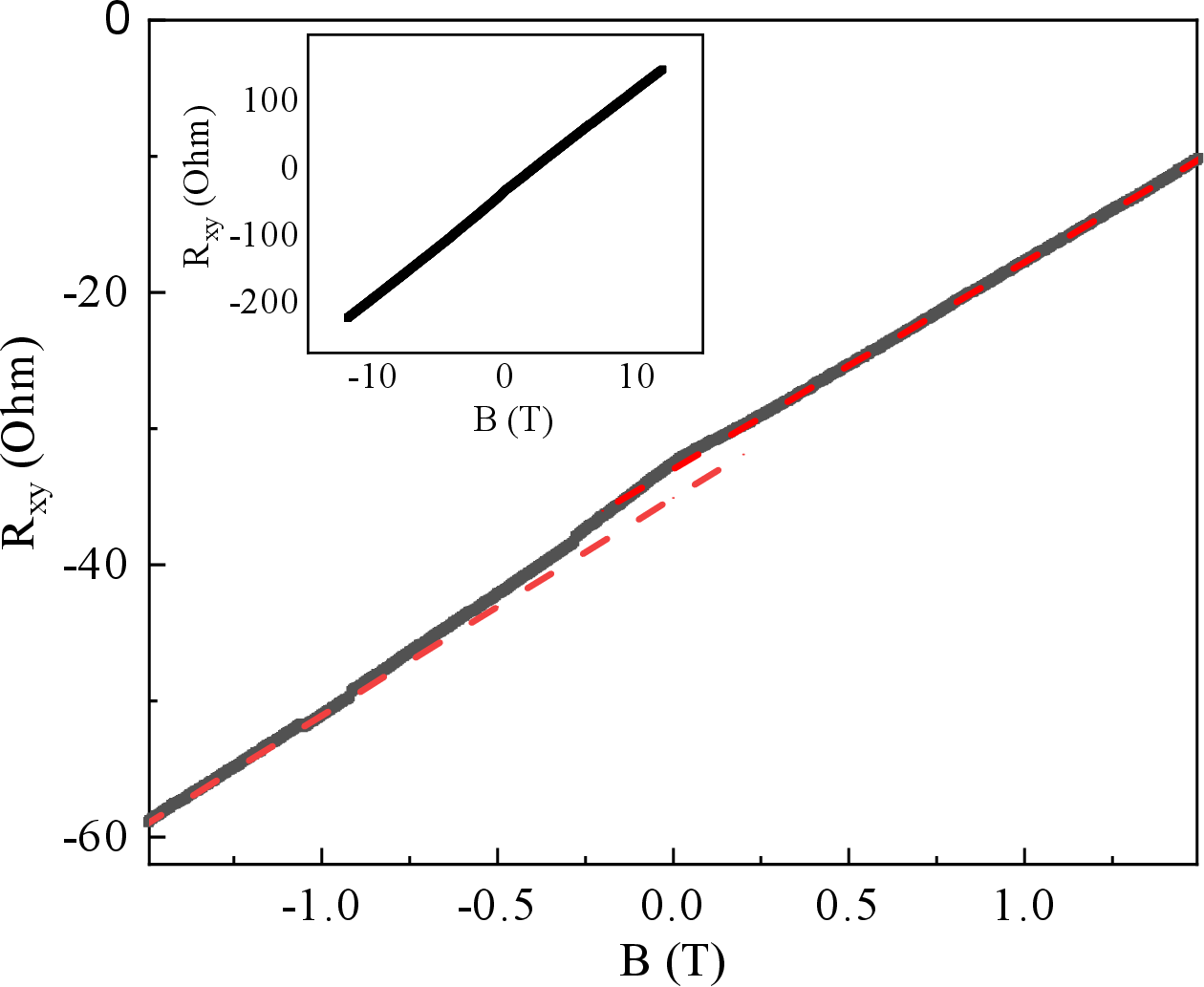}
	\caption{\label{fig:rawRxy} 
		Measured Hall resistance $R_{xy}$. Red dashed lines show asymptotics for high positive and negative fields. Inset shows $R_{xy}$ is large magnetic fields.  }
\end{figure}

In Fig.~\ref{fig:rawRxy} we present raw data of the Hall resistance for sample 36 measured at T = 1.64 K, prior to antisymmetrization procedure. The data contain inevitable admixture of the diagonal component of the resistivity tensor, i.e. magnetoresistance, i.e. even function of magnetic field. The Hall coefficient is obtained by considering only antisymmetric part of the data and its division over the magnetic field.
Nevertheless, even from this dataset it is easy to see that Hall slope at $B=0$ differs from the high-field asymptotics, that is a signature of the  Hall anomaly at low field.  Inset in Fig.~\ref{fig:rawRxy} shows Hall resistance in a wide range of magnetic field.

\section{Effect of quadratic and LMR correstion to HLN expression. }

\begin{figure}
	\centering
	\includegraphics[width=3.0in]{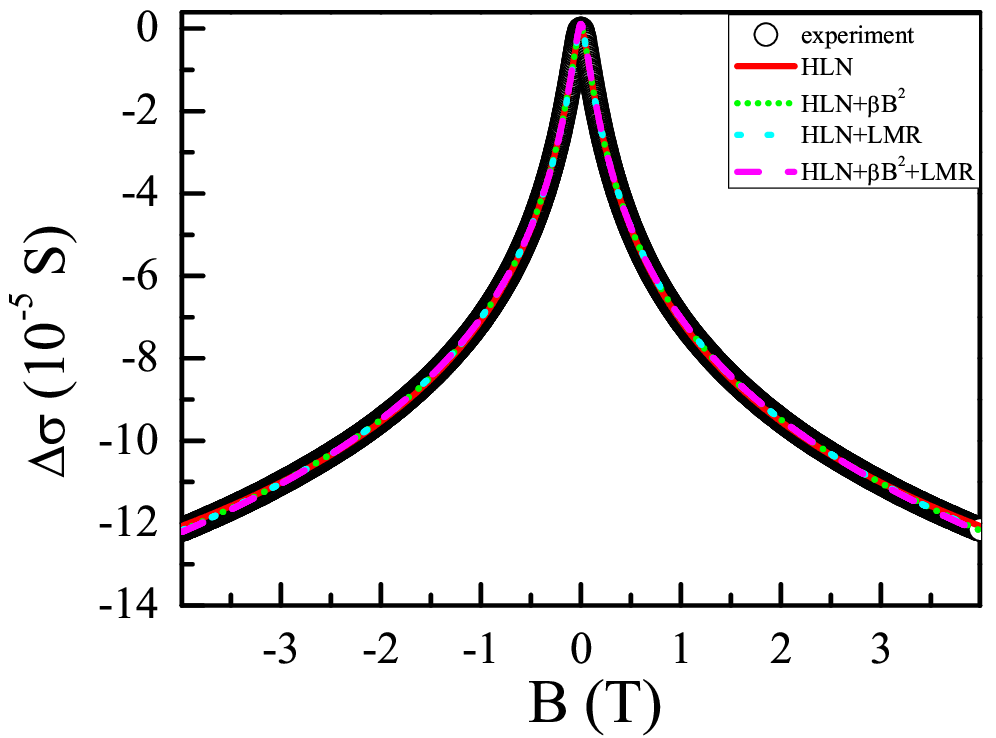}
	\caption{\label{fig:fitA} 
		Comparison of the magnetoresistance measured in sample 38 with Eq.~(\ref{HLN}) and Eqs.~(\ref{HLN-B}-\ref{HLN-LB}). }
\end{figure}

In the main text we show that the measured magnetoconductivity is described by HLN formula (\ref{HLN}) with unreasonably large $\alpha$ values. Sometimes~\cite{Assaf, singh22}, the elevated value of $\alpha$ in such a comparison is believed to be an artifact arising due to the quadratic contribution to magnetoconductivity $\delta \sigma \propto B^2$ or to linear magnetoresistance\cite{Parish} $\delta R \propto |B|$.
To rule out these explanation we fit our magnetoconductivity data with 
the following expressions
\begin{equation}\label{HLN-B}
\Delta\sigma(B) = \beta B^2 + \alpha \frac{e^2}{\pi h }[\Psi(\frac{B_0}{B} + \frac{1}{2}) - \ln(\frac{B_0}{B})]
\end{equation}
\begin{equation}\label{HLN-L}
\Delta\sigma(B) = \frac{1}{R_0 + R'|B|} + \alpha \frac{e^2}{\pi h }[\Psi(\frac{B_0}{B} + \frac{1}{2}) - \ln(\frac{B_0}{B})]
\end{equation}
\begin{multline}\label{HLN-LB}
\Delta\sigma(B) =\beta B^2 +  \frac{1}{R_0 + R'|B|} + \\
 \alpha \frac{e^2}{\pi h }[\Psi(\frac{B_0}{B} + \frac{1}{2}) - \ln(\frac{B_0}{B})]
\end{multline}
Here Eq.~(\ref{HLN-B}) corresponds to quadratic magnetoconductivity,  Eq.~(\ref{HLN-L}) to linear magnetoresistance and  Eq.~(\ref{HLN-LB})
includes both contributions. 
Fig.~\ref{fig:fitA} shows that all the equations can quantitatively describe the data. This result holds for all of our samples.

Table~\ref{Atabl} shows 
the values of $\alpha_{WAL}$ and $L_\varphi$ obtained from this fit with the least square method. Although $\alpha_{WAL}$ is 
actually slightly modified due to these contributions it stays larger (in absolute value) than unity for all the samples besides 45.  
Approximation of MC was also made for the smaller magnetic fields (1.5 T) (not shown in Table II), but the $\alpha_{WAL}$ value had not change much.

\begin{table}[ht]
	\centering
		\caption{\label{Atabl} Modifications of $\alpha$ and $L_\phi$ due to the quadratic and LMR contributions introduced in the model.   }
\begin{tabular}{|c|c|c|c|c|}
  \hline
  \, & \multicolumn{2}{c|}{$HLN$} &\multicolumn{2}{c|}{$HLN + B^2$} \\
  \cline{2-5}
  \, & $\alpha$ &  $L_{\varphi}$, nm & $\alpha$ & $L_{\varphi}$, nm \\
  \hline
  35 & -1.19 &72.2&-1.15&76.3  \\
  36 & -1.69 &72.4&-1.35&72.4  \\
  38 & -3.11 &81.3&-3.04&83.05 \\
  45 & -1.01 &58.2&-1   &59.7  \\
  46 & -2.65 &69.3&-2.47&71    \\
  \hline
  \, & \multicolumn{2}{c|}{$HLN+LMR$} &\multicolumn{2}{c|}{$HLN+LMR+B^2$} \\
  \cline{2-5}
  \, & $\alpha$ &  $L_{\varphi}$, nm & $\alpha$ & $L_{\varphi}$, nm \\
  \hline
  35 & -1.18&72.5&-1.17&74.2 \\
  36 & -1.72&69.7&-1.68&70.9 \\
  38 & -2.71&90.4&-2.85&87.3 \\
  45 & -0.86&65.7&-0.93&64.3 \\
  46 & -1.63&96.3&-1.76&88.8 \\
  \hline
\end{tabular}
\end{table}

\end{document}